# Surface and Bulk Effects of K in $Cu_{1-x}K_xIn_{1-y}Ga_ySe_2$ Solar Cells

Christopher P. Muzzillo[a,b] and Timothy J. Anderson[b]

[a]National Renewable Energy Laboratory, 15013 Denver West Pkwy, Golden, CO 80401, USA

[b]Department of Chemical Engineering, University of Florida, 1030 Center Dr, Gainesville, FL 32611, USA

Abstract

Two strategies for enhancing photovoltaic (PV) performance in chalcopyrite solar cells were investigated: $Cu_{1-x}K_xIn_{1-y}Ga_ySe_2$ absorbers with low K content (K/(K+Cu), or $x \sim 0.07$) distributed throughout the bulk, and $CuIn_{1-y}Ga_ySe_2$ absorbers with $KIn_{1-y}Ga_ySe_2$ grown on their surfaces. For the Ga-free case, increased temperature improved PV performance in the $KInSe_2$ surface absorbers, but not in the bulk $x \sim 0.07$ absorbers. Growth temperature also increased $KInSe_2$ phase fraction, relative to $Cu_{1-x}K_xInSe_2$ alloys—evidence that surface $KInSe_2$ improved performance more than bulk $KInSe_2$. Surface $KIn_{1-y}Ga_ySe_2$ and bulk $x \sim 0.07$ $Cu_{1-x}K_xIn_{1-y}Ga_ySe_2$ films with Ga/(Ga+In), or $y$ of 0.3 and 0.5 also had improved efficiency, open-circuit voltage ($V_{OC}$), and fill factor (FF), relative to $CuIn_{1-y}Ga_ySe_2$ baselines. On the other hand, $y \sim 1$ absorbers did not benefit from K introduction. Similar to $Cu_{1-x}K_xInSe_2$, the formation of $Cu_{1-x}K_xGaSe_2$ alloys was favored at low temperatures and high substrate Na content, relative to the formation of mixed-phase $CuGaSe_2 + KGaSe_2$. $KIn_{1-y}Ga_ySe_2$ alloys were grown for the first time, as evidenced by X-ray diffraction and ultraviolet/visible spectroscopy. For all Ga/(Ga+In) compositions, the surface $KIn_{1-y}Ga_ySe_2$ absorbers had superior PV performance in buffered and buffer-free devices. However, the bulk x ~ 0.07 absorbers only outperformed the baselines in buffered devices. The data demonstrate that $KIn_{1-y}Ga_ySe_2$ passivates the surface of $CuIn_{1-y}Ga_ySe_2$ to increase efficiency, $V_{OC}$, and FF, while bulk $Cu_{1-x}K_xIn_{1-y}Ga_ySe_2$ absorbers with x ~ 0.07 enhance efficiency, $V_{OC}$, and FF by some other mechanism.

Graphical abstract

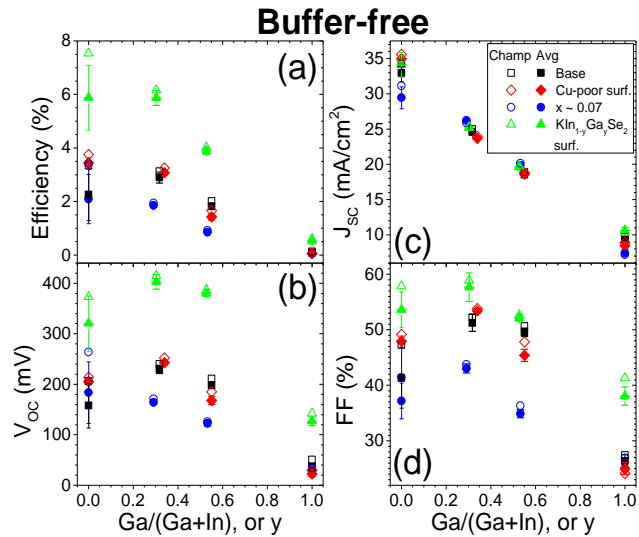

Highlights

- $Cu_{1-x}K_xIn_{1-y}Ga_ySe_2$ alloys formed at low temperature and high substrate surface Na

- $KIn_{1-y}Ga_ySe_2$ was grown for the first time

- Bulk $Cu_{1-x}K_xIn_{1-y}Ga_ySe_2$ with $x \sim 0.07$ had improved performance at $y \sim 0$, 0.3 and 0.5
- Surface $KIn_{1-y}Ga_ySe_2$ improved performance, but bulk $KIn_{1-y}Ga_ySe_2$ did not
- Bulk $x \sim 0.07$ and surface $KIn_{1-y}Ga_ySe_2$ improved performance by different mechanisms

1. Introduction

Recent reports have detailed photovoltaic (PV) power conversion efficiency enhancements when potassium fluoride and selenium were co-evaporated onto $Cu(In,Ga)(Se,S)_2$ (CIGS) absorbers at around 350°C (KF post-deposition treatment (PDT)) [1-21]. RbF has also been used [3, 17, 18]. Specifically, 6 of the last 8 world record CIGS efficiencies have employed a KF (or RbF) PDT [1, 3, 4, 12, 17, 22], advancing the record efficiency from 20.3 to 22.6% in just ~3.5 yr. KF PDT successes in the laboratory have now been extended to commercially-relevant chalcogenized CIGS absorbers [12], full size (0.75 m$^2$) modules [13], and Cd-free Zn(O,S) buffers [2, 12, 23]. Although the mechanisms responsible for these efficiency improvements are not clear, the KF PDT has been associated with multiple phenomena: *increased* hole concentration (e.g., by consuming $In_{Cu}$ compensating donors to produce $K_{Cu}$ neutral defects [24]) [5, 7, 8, 11, 14, 15, 19, 25-28], *decreased* hole concentration (by consuming $Na_{Cu}$ to produce $In_{Cu}$ compensating donors [1], or by forming $(K-K)_{Cu}$ dumbbell interstitial donors [29, 30]) [1, 8, 10, 16, 31], Na depletion or formation of soluble Na chemical(s) [1, 5, 7, 8, 10, 13, 14, 26, 27, 32, 33], Ga depletion at the surface [1, 8, 10, 13-15, 21, 32, 34, 35], Cu depletion at the surface [7, 13, 15, 18, 20, 21] resulting in better near-surface inversion

[1, 8, 10, 16, 36] or decreased valence band energy [14, 16, 21, 25, 34, 35, 37], grain boundary passivation [5, 12, 20, 31, 38], general defect passivation [2, 12, 27, 31], minority carrier lifetime enhancement [9, 15, 19], decreased minority carrier diffusion length [19], more Cd diffusion into the absorber (or a *less* abrupt interface) [1, 2, 8-10, 14, 25, 36, 39], a *more* abrupt absorber/buffer interface [16], morphology changes resulting in increased CdS nucleation sites [2, 10, 40], general changes in CdS growth [14, 34, 36, 41], formation of a passivating K-In-Ga-Se [10, 42], K-In-Se [14, 18, 34, 35, 41, 43], or In-Se [14, 18, 35, 41, 43] interfacial compound, formation of a current blocking interfacial compound [26], formation of elemental Se at the surface [14, 41, 43], formation of surface $Cu_{2-x}Se$ [14], $GaF_3$ [14], Ga-F [21], or In-F [21], increased formation of surface In-O and/or Ga-O after air exposure [14, 20, 25, 34, 37], consumption of a surface 'ordered vacancy compound' [14], decreased trap concentration [19, 32], reduced microscopic fluctuations in surface resistivity and potential [16, 31], and reduced nanoscopic fluctuations in potential [19]. Substrate surfaces with higher K content also led to CIGS with increased carrier concentrations [6, 24, 44], modified Cu-Ga-In interdiffusion [6, 44], as well as a K-rich and Cu-poor surface [45]. Similarly, an RbF PDT was associated with general defect passivation [3], and a CsF PDT caused Na and K depletion [17]. Another group co-evaporated KF, In, and Se onto CIGS, possibly forming $KInSe_2$ or K-doped amorphous $In_2Se_3$, with similar results to a KF PDT [41, 43]. A PDT without KF (just Se) has also been shown to significantly alter CIGS absorbers [46, 47], although most KF PDT results have not been compared to a Se PDT control sample. High absorber Na and K composition has also been linked to drastically accelerated degradation in PV performance [48], underscoring the importance of understanding alkali metal bonding in CIGS.

The ~29 phenomena associated with KF PDTs have obscured the underlying beneficial mechanisms, although it has been established that a relatively large amount of K is present at the p-n junction in the most efficient solar cells [1]. That observation motivated the study of K bonds in chalcopyrite-based material: $Cu_{1-x}K_xInSe_2$ alloys [49, 50], control of K bonds in Cu-K-In-Se material using substrate Na [33] and temperature [51], PV performance effects of K at the surface and bulk in $Cu_{1-x}K_xInSe_2$ [52], and PV performance [27] and surface spectroscopy [21] of bulk $Cu_{1-x}K_xIn_{1-y}Ga_ySe_2$ absorbers with K/(K+Cu), or x ~ 0.07 and Ga/(Ga+In), or y ~ 0.3. These studies on K bonding in chalcopyrites are presently extended to Ga alloys with y of 0 to 1, and connected with surface and bulk mechanisms for PV performance enhancement in $Cu_{1-x}K_xIn_{1-y}Ga_ySe_2$.

2. Methods

Most research on chalcopyrites utilizes three stage co-evaporation [1] or chalcogenization of metal precursors [12], but these two processes were presently avoided so that observations could not be attributed to altered cation profiles (K can affect cation diffusion [6, 44]). Co-evaporation of Cu, KF, In, Ga, and Se was performed on soda-lime glass (SLG) or SLG/Mo substrates at 400 – 600ºC. Films were 1 – 3 μm thick and had (Cu+K)/(In+Ga) of 0.85 and Se/(Cu+K+In+Ga) of 7 – 12, as measured by in situ electron impact emission spectroscopy, quartz crystal microbalance, and ex situ X-ray fluorescence (XRF). XRF was less certain due to K/In peak overlap [49], so in situ compositions were used in the present work unless noted. As previously discussed [49], phase-pure KInSe$_2$ was grown at a higher K/In of 2.12. Samples with more linear Tauc plot data were assumed to be more phase-pure, and so KIn$_{1-y}$Ga$_y$Se$_2$ with K/(Ga+In) of 1.96 and KGaSe$_2$ with K/Ga of 0.95 were compared in the present work. The slowest possible KF evaporation rates were used to improve control [49]. For KIn$_{1-y}$Ga$_y$Se$_2$ surface samples, the KF rate was ramped up and the Cu rate was ramped down over ~2 min near the end of the growth, followed by constant rate evaporation for ~2 min. Profilometry on the final film was used to infer individual layer thicknesses from in situ molar flux data, assuming constant density. The KIn$_{1-y}$Ga$_y$Se$_2$ surface process optimized for Ga/(Ga+In), or y of 0 [52] was used at all other Ga/(Ga+In) compositions (132 nm intermediate layer and a 75 nm surface layer with K/(K+Cu), or x ~ 0.41). Control samples with Cu-poor surfaces were prepared by using that same cation profile with no KF added. For comparison, the bulk x ~ 0.07 and KIn$_{1-y}$Ga$_y$Se$_2$ surface absorbers had 1.5 and 0.4 at.-% K by in situ measurement, respectively. Some absorbers were annealed with Se over-pressure after growth and without breaking vacuum. X-ray diffraction (XRD), XRF, secondary ion mass spectrometry (SIMS), and UV/visible spectroscopy were performed using previously described conditions [49, 51]. Unlike typical KF PDT procedures [1, 5-10, 14-16, 19, 25, 27, 34, 35, 37], absorbers were *not* rinsed before chemical bath deposition (CBD). Solar cells were fabricated with 50 nm CdS, 100 nm i-ZnO, 120 nm Al:ZnO, 50 nm Ni, and 3 µm Al, as previously detailed [53]. JV and QE were performed under formerly reported conditions [54].

3. Results

3.1. Solar cells

During co-evaporation of Cu-K-In-Se, substrate temperature was shown to determine the extent of $Cu_{1-x}K_xInSe_2$ alloy formation, relative to $CuInSe_2$ + $KInSe_2$ mixed-phases [51]. Solar cells made from bulk x ~ 0.07 CKIS alloy absorbers had excellent PV performance [52]. Additionally, $CuInSe_2$/$KInSe_2$ ($KInSe_2$ surface) absorbers were highly efficient [52]. In this study, growth temperature was varied during baseline, bulk x ~ 0.07 $Cu_{1-x}K_xInSe_2$ alloy, and $KInSe_2$ surface absorber growth processes. Absorbers were either grown at 400°C, grown at 400°C and immediately annealed to 600°C for 10 min, grown at 500°C, or grown at 600°C (Fig. 1). The baseline 500°C $CuInSe_2$ absorbers had higher $J_{SC}$ and lower FF, relative to 600°C, that was previously attributed to enhanced nucleation of CdS during CBD [52]. Relative to baseline changes from 500 to 600°C, the $KInSe_2$ surface had reproducibly greater efficiency, $V_{OC}$, and FF. On the other hand, the bulk x ~ 0.07 $Cu_{1-x}K_xInSe_2$ had lower efficiency, $V_{OC}$, $J_{SC}$, and FF on moving from 500 to 600°C (relative to the baseline), although the small number of samples made this conclusion less certain.

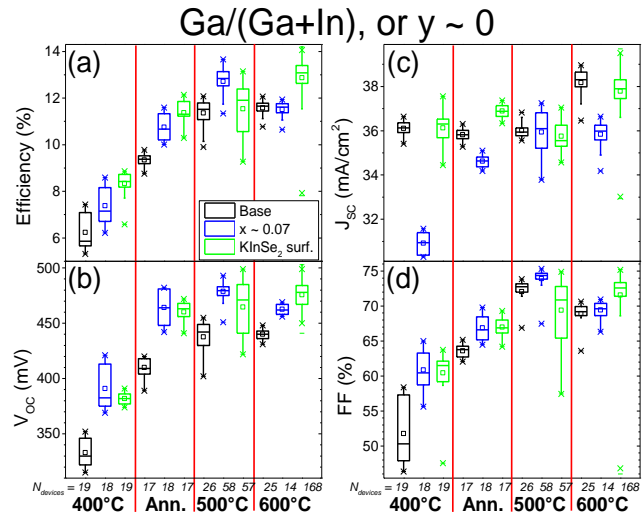

Fig. 1. Baseline (black), bulk x ~ 0.07 (blue), and $KInSe_2$ surface (green) efficiency (a), open-circuit voltage (b), short-circuit current density (c), and fill factor (d) at Ga/(Ga+In),

or y of 0 and grown at 400°C, grown at 400°C and annealed to 600°C, grown at 500°C, and grown at 600°C. Boxes are quartiles, crosses are 1st and 99th percentiles, squares are means, dashes are minima and maxima, and sample sizes ($N_{devices}$) are below the plots.

Next, the bulk x ~ 0.07 $Cu_{1-x}K_xInSe_2$ and $KInSe_2$ surface growth techniques were extended to $Cu_{1-x}K_xIn_{1-y}Ga_ySe_2$ alloys with Ga/(Ga+In), or y of 0.3, 0.5 and 1. Additional Cu-poor surface control absorbers were studied—they had the same cation profiles as the $KIn_{1-y}Ga_ySe_2$ surface absorbers, except no KF was added. For Ga/(Ga+In) ~ 0.3, the bulk x ~ 0.07 absorbers had substantially greater efficiency, $V_{OC}$, $J_{SC}$, and FF, relative to the $KIn_{1-y}Ga_ySe_2$ surfaces (Fig. 2). However, a low number of bulk x ~ 0.07 devices were studied in Fig. 2, so this finding should be verified in future work.

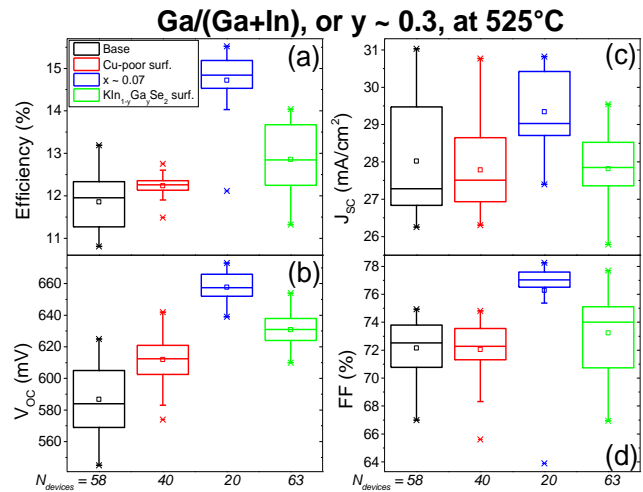

Fig. 2. Baseline (black), Cu-poor surface (red), bulk x ~ 0.07 (blue), and $KIn_{1-y}Ga_ySe_2$ surface (green) efficiency (a), open-circuit voltage (b), short-circuit current density (c), and fill factor (d) at Ga/(Ga+In), or y of 0.3 and grown at 525°C. Boxes are quartiles, crosses are 1st and 99th percentiles, squares are means, dashes are minima and maxima, and sample sizes ($N_{devices}$) are below the plots.

At higher Ga/(Ga+In) of 0.5, the efficiency increased from baseline, to Cu-poor surface, to bulk $x \sim 0.07$, and to $KIn_{1-y}Ga_ySe_2$ surface absorbers (Fig. 3). The $KIn_{1-y}Ga_ySe_2$ surface had remarkably higher $V_{OC}$ and slightly higher FF than bulk $x \sim 0.07$. The bulk $x \sim 0.07$ had greater $J_{SC}$ than the $KIn_{1-y}Ga_ySe_2$ surface, and this was due to increased long wavelength collection by QE (not shown). For Ga/(Ga+In) of 1, PV performance was poor, and there were no trends among the different absorber processes (Fig. 4).

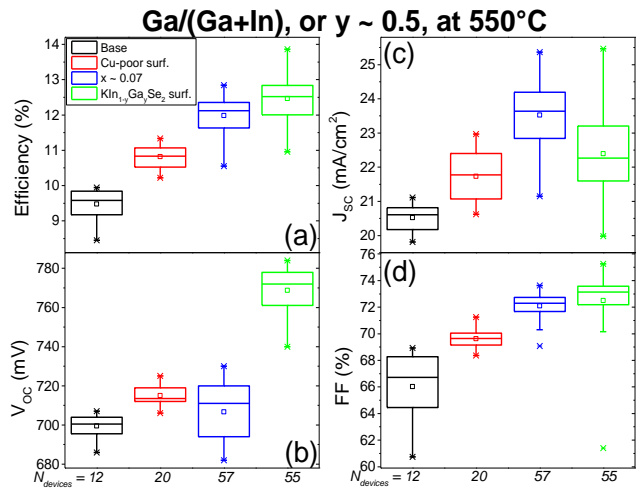

Fig. 3. Baseline (black), Cu-poor surface (red), bulk $x \sim 0.07$ (blue), and $KIn_{1-y}Ga_ySe_2$ surface (green) efficiency (a), open-circuit voltage (b), short-circuit current density (c), and fill factor (d) at Ga/(Ga+In), or y of 0.5 and grown at 550°C. Boxes are quartiles, crosses are 1st and 99th percentiles, squares are means, dashes are minima and maxima, and sample sizes ($N_{devices}$) are below the plots.

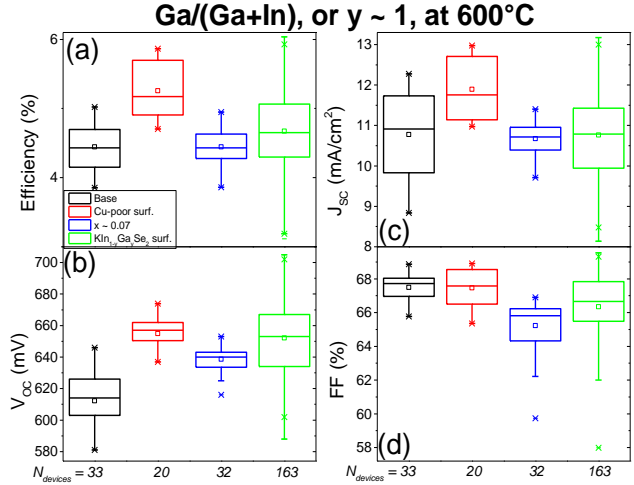

Fig. 4. Baseline (black), Cu-poor surface (red), bulk x ~ 0.07 (blue), and $KIn_{1-y}Ga_ySe_2$ surface (green) efficiency (a), open-circuit voltage (b), short-circuit current density (c), and fill factor (d) at Ga/(Ga+In), or y of 1 and grown at 600°C. Boxes are quartiles, crosses are 1st and 99th percentiles, squares are means, dashes are minima and maxima, and sample sizes ($N_{devices}$) are below the plots.

The PV performance of 'buffer-free' ($Cu_{1-x}K_xIn_{1-y}Ga_ySe_2$/i-ZnO/Al:ZnO) devices was measured for each absorber growth technique and each Ga/(Ga+In) composition (Fig. 5). Despite the different growth temperatures used to construct Fig. 5, as well as the relatively poor performance, a clear trend emerged: $KIn_{1-y}Ga_ySe_2$ surfaces had greater efficiency, $V_{OC}$, and FF, relative to the baselines. On the other hand, bulk x ~ 0.07 absorbers had similar or worse efficiency, $V_{OC}$, and FF, relative to the baselines. These trends held for all Ga/(Ga+In) compositions.

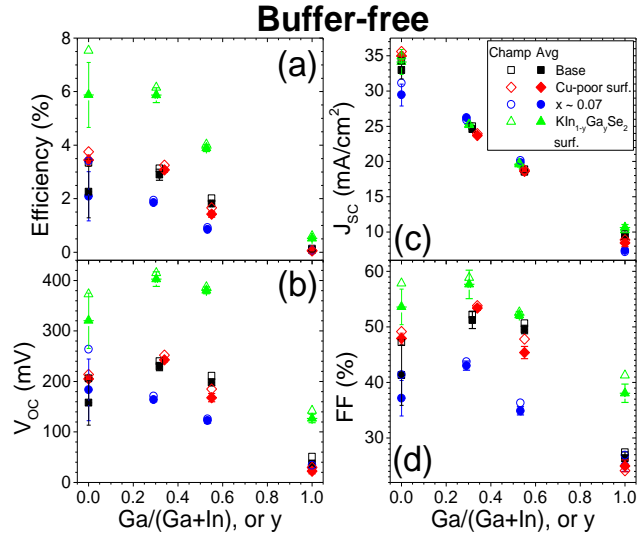

Fig. 5. Buffer-free champion (empty symbols) and average (filled symbols) efficiency (a), open-circuit voltage (b), short circuit current density (c), and fill factor (d) for baseline (black squares), Cu-poor surface (red diamonds), bulk $x \sim 0.07$ (blue circles), and $KIn_{1-y}Ga_ySe_2$ surface (green triangles) absorbers as a function of Ga/(Ga+In), or y composition by XRF. Growth temperatures are varied, error bars are standard deviations, and the number of devices was 59, 27, 25, and 27 for Ga/(Ga+In) of 0, 0.3, 0.5, and 1, respectively.

3.2. Phase growth

$Cu_{1-x}K_xInSe_2$ alloy formation was previously shown to drive Na diffusion out from the substrate [27, 33]. This was also observed for Ga/(Ga+In) of 0.3, 0.5 and 1 in the present study (SIMS in Fig. 6, 7 and 8). As expected, the bulk $x \sim 0.07$ films had increased K throughout, while the $KIn_{1-y}Ga_ySe_2$ surfaces had increased K near the surface for all Ga/(Ga+In) compositions. The Cu/(Ga+In) profiles were similar for the baseline, Cu-poor surface, and bulk $x \sim 0.07$ absorbers, so the Cu-poor surface process may not have actually confined Cu-poor material to the surface (Fig. 9, 10 and 11). As expected from the cation flux profiles, the $KIn_{1-y}Ga_ySe_2$ surfaces had reduced Cu/(Ga+In) compositions near the surface, relative to all other films. However, the $KIn_{1-y}Ga_ySe_2$ surfaces also exhibited reduced Ga/(Ga+In) compositions near the surface (Fig. 9 and 10). There are multiple explanations for this observation: unintentional Ga/(Ga+In) flux alteration at the end of absorber growth, greater chemical affinity of K-Se and In-Se (relative to K-Se and Ga-Se), faster kinetics of K-Se and In-Se incorporation (relative to K-Se and Ga-Se), or different In and Ga SIMS sputtering yields for $KIn_{1-y}Ga_ySe_2$ (relative to $CuIn_{1-y}Ga_ySe_2$). More study will be needed to affirm or refute each possibility.

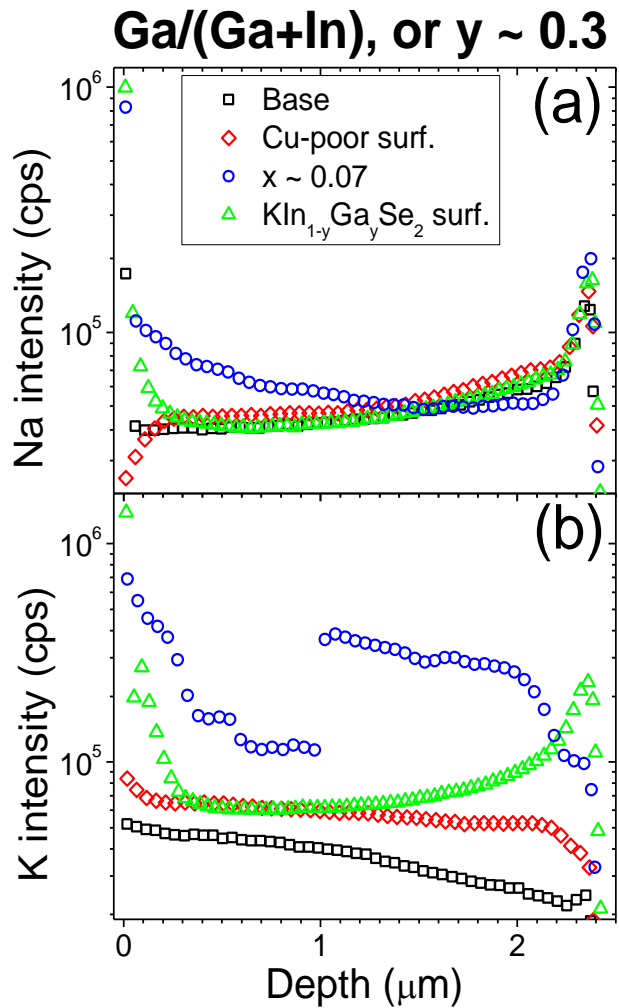

Fig. 6. SIMS Na (a) and K (b) profiles for baseline (black squares), Cu-poor surface (red diamonds), bulk x ~ 0.07 (blue circles), and KIn$_{1-y}$Ga$_y$Se$_2$ surface (green triangles) absorbers with Ga/(Ga+In), or y of 0.3 (grown at 525°C). A depth of 0 is the absorbers' free surface; data is scaled so all films are equally thick.

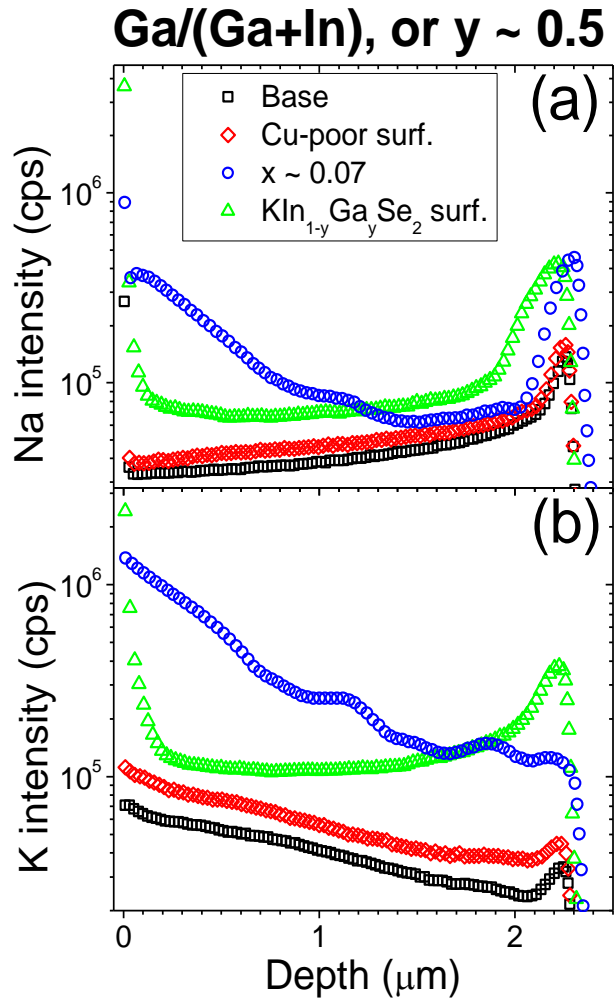

Fig. 7. SIMS Na (a) and K (b) profiles for baseline (black squares), Cu-poor surface (red diamonds), bulk x ~ 0.07 (blue circles), and KIn$_{1-y}$Ga$_y$Se$_2$ surface (green triangles) absorbers with Ga/(Ga+In), or y of 0.5 (grown at 550°C). A depth of 0 is the absorbers' free surface; data is scaled so all films are equally thick.

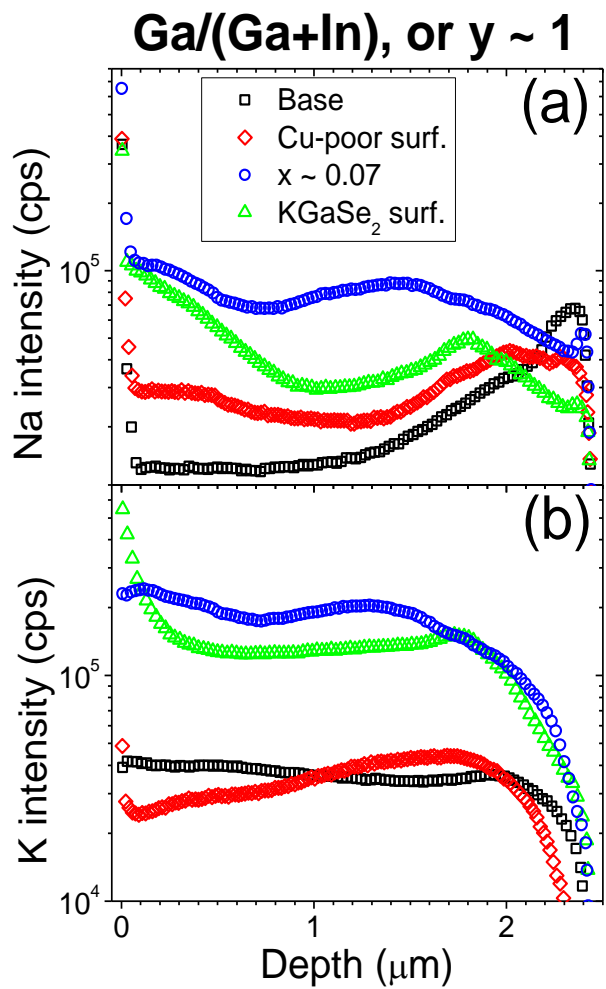

Fig. 8. SIMS Na (a) and K (b) profiles for baseline (black squares), Cu-poor surface (red diamonds), bulk x ~ 0.07 (blue circles), and KGaSe$_2$ surface (green triangles) absorbers with Ga/(Ga+In), or y of 1 (grown at 600°C). A depth of 0 is the absorbers' free surface; data is scaled so all films are equally thick.

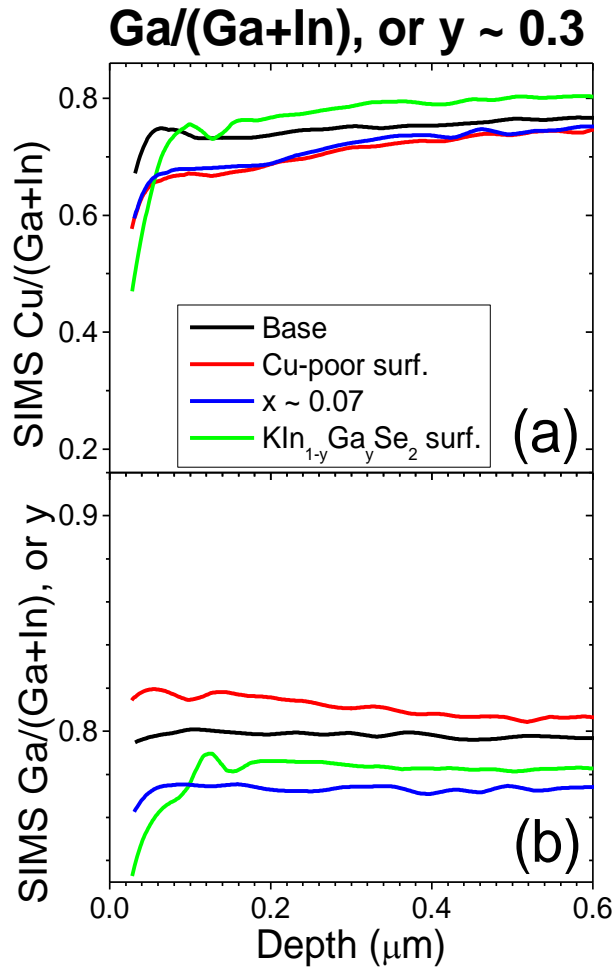

Fig. 9. SIMS Cu/(Ga+In) (a) and Ga/(Ga+In) (b) profiles for baseline (black), Cu-poor surface (red), bulk $x \sim 0.07$ (blue), and $KIn_{1-y}Ga_ySe_2$ surface (green) absorbers with Ga/(Ga+In), or y of 0.3 (grown at 525°C). A depth of 0 is the absorbers' free surface; data is scaled so all films are equally thick; $CuCs^+$, $GaCs^+$ and $InCs^+$ were used.

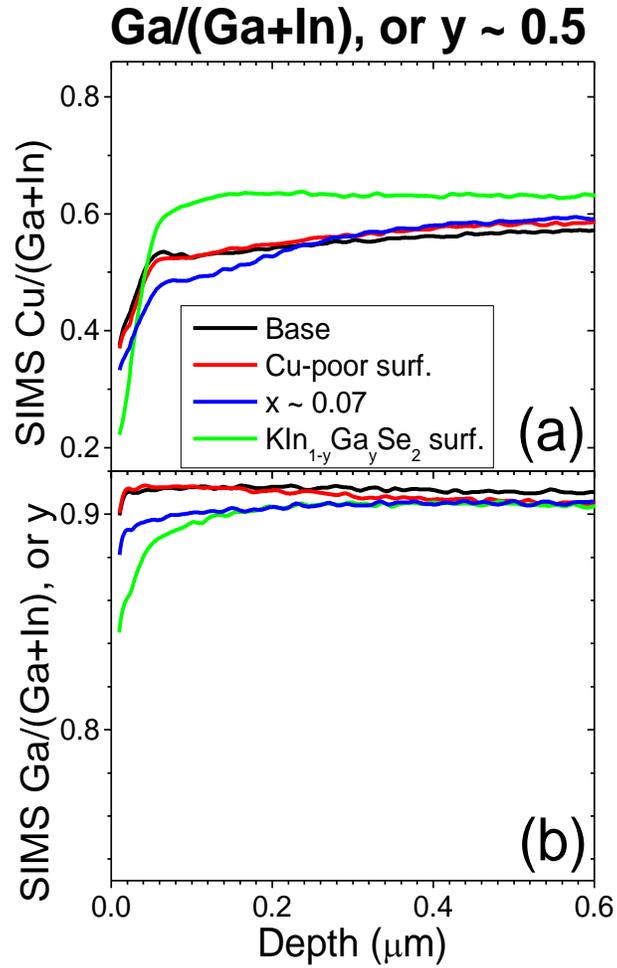

Fig. 10. SIMS Cu/(Ga+In) (a) and Ga/(Ga+In) (b) profiles for baseline (black), Cu-poor surface (red), bulk $x \sim 0.07$ (blue), and $KIn_{1-y}Ga_ySe_2$ surface (green) absorbers with Ga/(Ga+In), or y of 0.5 (grown at 550°C). A depth of 0 is the absorbers' free surface; data is scaled so all films are equally thick; $CuCs^+$, $GaCs^+$ and $InCs^+$ were used.

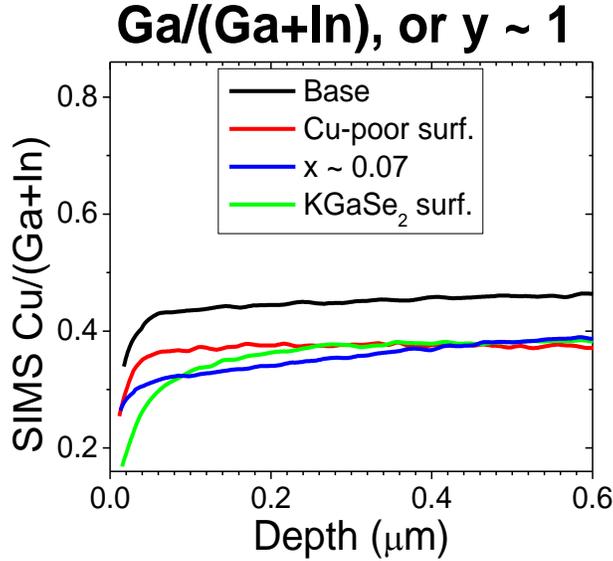

Fig. 11. SIMS Cu/(Ga+In) profiles for baseline (black), Cu-poor surface (red), bulk x ~ 0.07 (blue), and KGaSe$_2$ surface (green) absorbers with Ga/(Ga+In), or y of 1 (grown at 600°C). A depth of 0 is the absorbers' free surface; data is scaled so all films are equally thick; CuCs$^+$ and GaCs$^+$ were used.

The extent to which Cu$_{1-x}$K$_x$GaSe$_2$ alloys formed was examined by growing on SLG and SLG/Mo substrates at 400°C and 500°C. XRD scans are shown for x of 0 and 0.5 in Fig. 12 for the high Na substrate (SLG) at low growth temperature (400°C). The x ~ 0 sample had a peak near 26.2° 2θ that was tentatively labeled Cu$_2$Se, which can form at reduced growth temperatures even for Cu-poor compositions (Cu/Ga < 1) [55, 56]. The peak could also relate to stacking faults (similar to CuInSe$_2$ [57]), and more study will be needed to support either origin. The x ~ 0.5 film had chalcopyrite peaks shifted to smaller d-spacing, as well as small KGaSe$_2$ peaks [58]. These peak shifts were not observed on Mo substrates or at higher growth temperature—taken as evidence that Cu$_{1-x}$K$_x$GaSe$_2$ alloy formation is favored at low temperature and high Na substrates, similar to Cu$_{1-}$

$_x$K$_x$InSe$_2$ alloy formation [33, 51]. UV/visible spectroscopy revealed a shifted absorption onset from x ~ 0 to 0.5 (Fig. 13). The extrapolated band gap was shifted by 0.18 eV (Table 1), and taken as further evidence of Cu$_{1-x}$K$_x$GaSe$_2$ alloy formation. Previous experimental and theoretical studies found some evidence of Cu$_{1-x}$Na$_x$In$_{1-y}$Ga$_y$Se$_2$ compounds [59-64], and future work comparing Na and K alloys could be mutually informative.

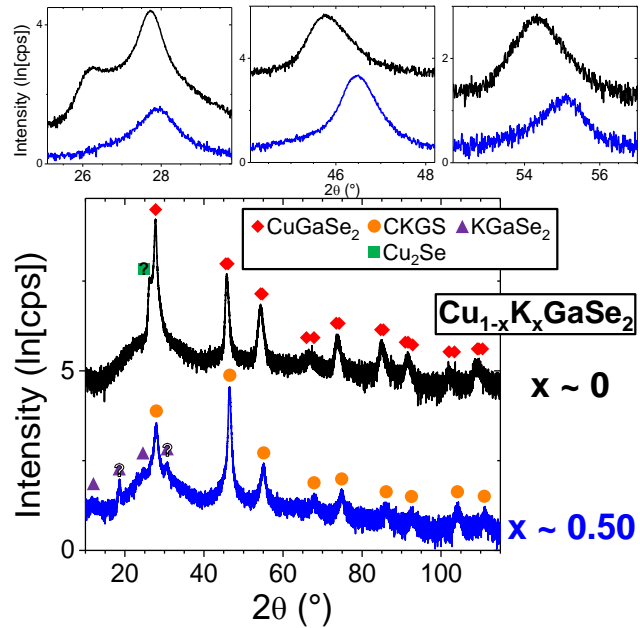

Fig. 12. XRD scans of Cu$_{1-x}$K$_x$GaSe$_2$ (Ga/(Ga+In), or y of 1) films with K/(K+Cu), or x of 0 (top; black) and 0.50 (bottom; blue) grown at 400°C on SLG. CuGaSe$_2$, Cu$_{1-x}$K$_x$GaSe$_2$, KGaSe$_2$, and Cu$_2$Se peaks are labeled with red diamonds, orange circles, purple triangles, and green squares, respectively.

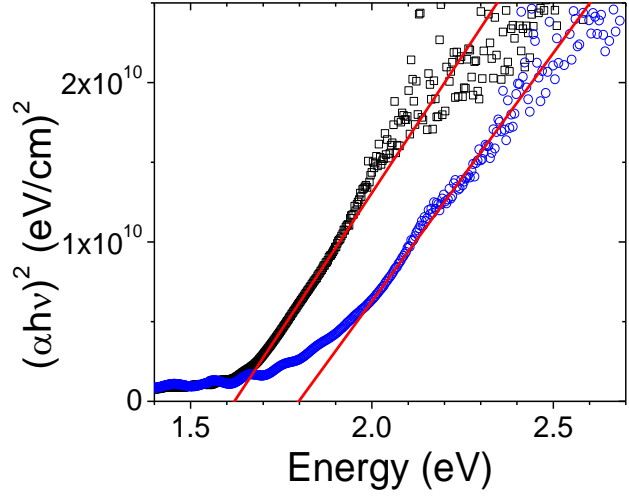

Fig. 13. Tauc plot (the product of absorption coefficient and photon energy squared, $(\alpha h\nu)^2$, versus photon energy, $h\nu$) for $Cu_{1-x}K_xGaSe_2$ (Ga/(Ga+In), or y of 1) films with K/(K+Cu), or x of 0 (black squares) and 0.50 (blue circles). Red lines are least squares fits to the data.

Table 1. Band gaps of $Cu_{1-x}K_xIn_{1-y}Ga_ySe_2$ compounds with varied compositions. Detailed $Cu_{1-x}K_xInSe_2$ band gaps were previously published [49].

| Compound | K/(K+Cu), or x | Ga/(Ga+In), or y | Band gap (eV) | Reference |
|---|---|---|---|---|
| $CuGaSe_2$ | 0 | 1 | 1.62(1) | This work |
|  |  |  | 1.648(2) | [65] |
| $Cu_{0.50}K_{0.50}GaSe_2$ | 0.50 | 1 | 1.80(5) | This work |
| $Cu_{0.33}K_{0.67}GaSe_2$ | 0.67 | 1 | 1.72 | [66] |
| $KGaSe_2$ | 1 | 1 | 3.22(1) | This work |
|  |  |  | 2.60(2) | [67] |
| $KIn_{0.43}Ga_{0.57}Se_2$ | 1 | 0.57 | 2.77(1) | This work |
| $KInSe_2$ | 1 | 0 | 2.71(1) | [49] |
|  |  |  | 2.68 | [68] |

The published structures of $KInSe_2$ [69] and $KGaSe_2$ [58] were in good agreement with their XRD patterns in Fig. 14. A film with Ga/(Ga+In) of 0.57 was also grown, and its XRD peaks were between those of $KInSe_2$ and $KGaSe_2$—evidence of $KIn_{1-y}Ga_ySe_2$ alloy formation in Fig. 14. The $KIn_{1-y}Ga_ySe_2$ film also had a band gap of 2.77 eV, which fell between that of $KInSe_2$ and $KGaSe_2$, and was taken as further evidence of alloying (Fig. 15 and Table 1). The band gap of $KGaSe_2$ was 3.22 eV (Fig. 15 and Table 1), substantially greater than the value reported for bulk crystals (2.60 eV [67]). It is speculated that this difference could relate to a compositional homogeneity range in $KGaSe_2$, or some impurity phase ($K_5GaSe_4$ and $K_3GaSe_3$ have been reported [70, 71]).

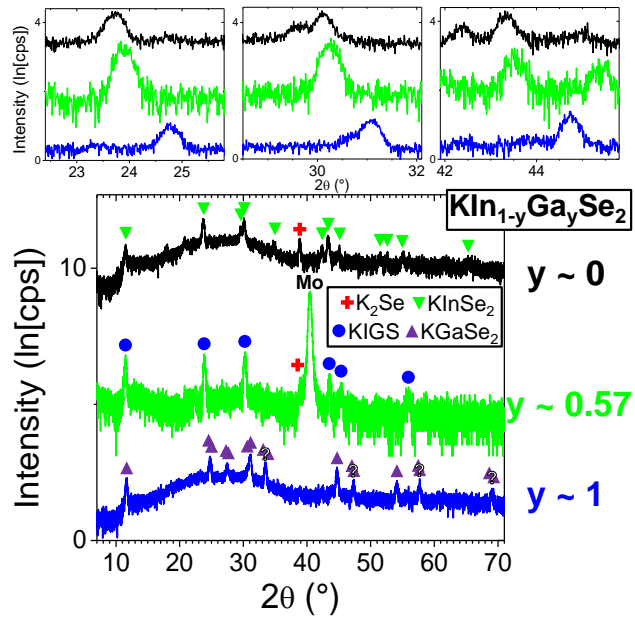

Fig. 14. XRD scans of $KIn_{1-y}Ga_ySe_2$ (K/(K+Cu), or x of 1) films with Ga/(Ga+In), or y of 0 (top; black), 0.57 (middle; green), and 1 (bottom; blue) grown at 500°C grown on Mo or SLG. Mo, $K_2Se$, $KInSe_2$, $KIn_{1-y}Ga_ySe_2$, and $KGaSe_2$ peaks are labeled with 'Mo,' red plusses, green down triangles, blue circles, and purple up triangles, respectively.

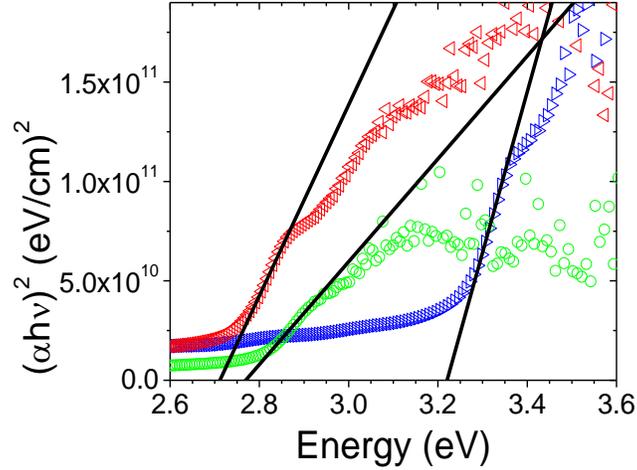

Fig. 15. Tauc plot (the product of absorption coefficient and photon energy squared, $(\alpha h\nu)^2$, versus photon energy, $h\nu$) for $KIn_{1-y}Ga_ySe_2$ (K/(K+Cu), or x of 1) films with Ga/(Ga+In), or y of 0 (red left-pointing triangles), 0.57 (green circles), and 1 (blue right-pointing triangles). Black lines are least squares fits to the data.

4. Discussion

Growth temperature was found to determine $Cu_{1-x}K_xInSe_2$ alloy formation, relative to $CuInSe_2$ + $KInSe_2$ formation [51]. The $KInSe_2$ surface had higher efficiency, $V_{OC}$, and FF at 600°C, relative to 500°C. Therefore, differences in $KInSe_2$ surface phase fraction significantly altered PV performance. Increasing substrate temperature for bulk x ~ 0.07, and hence bulk $KInSe_2$ phase fraction, appeared to detriment performance. This affirmed that a greater phase fraction of $KInSe_2$ could be tolerated at the surface (relative to the bulk) in efficient devices, as previously shown [52]. It may also be evidence that K can improve performance by different mechanisms at the surface and bulk [52].

In this work, substrate surface Na content and temperature were found to determine $Cu_{1-x}K_xGaSe_2$ alloy formation, relative to $CuGaSe_2$ + $KGaSe_2$ mixed-phases

(Fig. 12 and 13). The enhanced performance of bulk $x \sim 0.07$, relative to $KIn_{1-y}Ga_ySe_2$ surfaces at Ga/(Ga+In) of 0.3 and 525°C could have been affected by the Ga/(Ga+In) composition, or by the relatively low growth temperature. Likewise, the $KIn_{1-y}Ga_ySe_2$ surface performed better than the bulk $x \sim 0.07$ at Ga/(Ga+In) of 0.5 and 550°C, which could have been affected by Ga/(Ga+In), or those growths' increased temperatures. The temperature and Ga/(Ga+In) effects should be isolated in future experiments.

Further evidence that surface and bulk K can improve PV performance by different mechanisms was found in the buffer-free results. The $KIn_{1-y}Ga_ySe_2$ surface absorbers had enhanced performance in buffered and buffer-free devices, while the bulk $x \sim 0.07$ absorbers only had enhanced performance in buffered devices. This was evidence that $KIn_{1-y}Ga_ySe_2$ reduced recombination at the absorber/ZnO interface—possibly by protecting the underlying absorber from oxidation or sputtering damage, or by passivating the absorber. Previous work associated $KInSe_2$ surfaces with increased minority carrier lifetimes and inversion near the surface [52]. Together, these data suggest that $KIn_{1-y}Ga_ySe_2$ passivates the surface of $CuIn_{1-y}Ga_ySe_2$, while the bulk $x \sim 0.07$ $Cu_{1-x}K_xIn_{1-y}Ga_ySe_2$ absorbers improve performance by some other mechanism.

5. Conclusions

Increasing substrate temperature was previously found to increase $KInSe_2$ phase fraction during Cu-K-In-Se growth [51]. In this study, increased temperature improved PV performance of $KInSe_2$ surface absorbers, but not bulk $x \sim 0.07$ absorbers—further evidence that K can improve performance by different mechanisms at the surface and in the bulk absorber [52]. Both mechanisms also benefitted efficiency, $V_{OC}$, and FF in $Cu_{1-}$

$_x$K$_x$In$_{1-y}$Ga$_y$Se$_2$ films with Ga/(Ga+In), or y of 0.3 and 0.5, but not for y of 1. Similar to the Cu$_{1-x}$K$_x$InSe$_2$ system [33, 49, 51], the formation of Cu$_{1-x}$K$_x$GaSe$_2$ alloys was favored at low temperatures and high substrate Na content, relative to the formation of mixed-phase CuGaSe$_2$ + KGaSe$_2$. KIn$_{1-y}$Ga$_y$Se$_2$ alloys were grown for the first time, as evidenced by XRD and UV/visible spectroscopy. Surface KIn$_{1-y}$Ga$_y$Se$_2$ absorbers had superior PV performance in buffered and buffer-free devices, while bulk x ~ 0.07 absorbers only outperformed the baselines in buffered devices. The study has shown that KIn$_{1-y}$Ga$_y$Se$_2$ passivates the surface of CuIn$_{1-y}$Ga$_y$Se$_2$ to increase efficiency, $V_{OC}$, and FF, while bulk Cu$_{1-x}$K$_x$In$_{1-y}$Ga$_y$Se$_2$ absorbers with x ~ 0.07 enhance efficiency, $V_{OC}$, and FF by some other mechanism.

Acknowledgements

The authors thank Stephen Glynn, Lorelle Mansfield, and Carolyn Beall for assistance with experiments, Clay DeHart for contact deposition, Karen Bowers for device processing, and Matt Young for SIMS. This work was supported by the U.S. Department of Energy under Contract DE-AC36-08GO28308 with the National Renewable Energy Laboratory.

References

[1] A. Chirilă, P. Reinhard, F. Pianezzi, P. Bloesch, A.R. Uhl, C. Fella, L. Kranz, D. Keller, C. Gretener, H. Hagendorfer, D. Jaeger, R. Erni, S. Nishiwaki, S. Buecheler, A.N. Tiwari, Potassium-induced surface modification of Cu(In,Ga)Se$_2$ thin films for high-efficiency solar cells, Nat Mater, 12 (2013) 1107-1111.
[2] T.M. Friedlmeier, P. Jackson, A. Bauer, D. Hariskos, O. Kiowski, R. Wuerz, M. Powalla, Improved Photocurrent in Cu(In,Ga)Se$_2$ Solar Cells: From 20.8% to 21.7%

Efficiency with CdS Buffer and 21.0% Cd-Free, Photovoltaics, IEEE Journal of, 5 (2015) 1487-1491.
[3] P. Jackson, D. Hariskos, R. Wuerz, O. Kiowski, A. Bauer, T.M. Friedlmeier, M. Powalla, Properties of Cu(In,Ga)Se$_2$ solar cells with new record efficiencies up to 21.7%, physica status solidi (RRL) – Rapid Research Letters, 9 (2015) 28-31.
[4] P. Jackson, D. Hariskos, R. Wuerz, W. Wischmann, M. Powalla, Compositional investigation of potassium doped Cu(In,Ga)Se$_2$ solar cells with efficiencies up to 20.8%, physica status solidi (RRL) – Rapid Research Letters, 8 (2014) 219-222.
[5] A. Laemmle, R. Wuerz, M. Powalla, Efficiency enhancement of Cu(In,Ga)Se$_2$ thin-film solar cells by a post-deposition treatment with potassium fluoride, physica status solidi (RRL) – Rapid Research Letters, 7 (2013) 631-634.
[6] A. Laemmle, R. Wuerz, M. Powalla, Investigation of the effect of potassium on Cu(In,Ga)Se$_2$ layers and solar cells, Thin Solid Films, 582 (2015) 27-30.
[7] L.M. Mansfield, R. Noufi, C.P. Muzzillo, C. DeHart, K. Bowers, B. To, J.W. Pankow, R.C. Reedy, K. Ramanathan, Enhanced performance in Cu(In,Ga)Se$_2$ solar cells fabricated by the two-step selenization process with a potassium fluoride postdeposition treatment, Photovoltaics, IEEE Journal of, 4 (2014) 1650-1654.
[8] F. Pianezzi, P. Reinhard, A. Chirilă, B. Bissig, S. Nishiwaki, S. Buecheler, A.N. Tiwari, Unveiling the effects of post-deposition treatment with different alkaline elements on the electronic properties of CIGS thin film solar cells, Phys. Chem. Chem. Phys., 16 (2014) 8843-8851.
[9] F. Pianezzi, P. Reinhard, A. Chirilă, S. Nishiwaki, B. Bissig, S. Buecheler, A.N. Tiwari, Defect formation in Cu(In,Ga)Se$_2$ thin films due to the presence of potassium during growth by low temperature co-evaporation process, J. Appl. Phys., 114 (2013) 194508-194508.
[10] P. Reinhard, B. Bissig, F. Pianezzi, E. Avancini, H. Hagendorfer, D. Keller, P. Fuchs, M. Döbeli, C. Vigo, P. Crivelli, S. Nishiwaki, S. Buecheler, A.N. Tiwari, Features of KF and NaF Postdeposition Treatments of Cu(In,Ga)Se$_2$ Absorbers for High Efficiency Thin Film Solar Cells, Chem. Mater., 27 (2015) 5755-5764.
[11] J.M. Raguse, C.P. Muzzillo, J.R. Sites, Effects of sodium and potassium on the photovoltaic performance of CIGS solar cells, Journal of Photovoltaics, (2016) submitted.
[12] R. Kamada, T. Yagioka, S. Adachi, A. Handa, K.F. Tai, T. Kato, H. Sugimoto, New world record Cu(In,Ga)(Se,S)$_2$ thin film solar cell efficiency beyond 22%, in: Photovoltaic Specialist Conference (PVSC), 2016 IEEE 43rd, Portland, OR, 2016, pp. in press.
[13] O. Lundberg, E. Wallin, V. Gusak, S. Södergren, S. Chen, S. Lotfi, F. Chalvet, U. Malm, N. Kaihovirta, P. Mende, G. Jaschke, P. Kratzert, J. Joel, M. Skupinski, P. Lindberg, T. Jarmar, J. Lundberg, J. Mathiasson, L. Stolt, Improved CIGS Modules by KF Post Deposition Treatment and Reduced Cell-to-Module Losses, in: 43rd IEEE Photovoltaic Specialists Conference, Portland, OR, 2016, pp. 1-4.
[14] T. Lepetit, Influence of KF post deposition treatment on the polycrystalline Cu(In,Ga)Se$_2$/CdS heterojunction formation for photovoltaic application, in: Mater. Sci., Université de Nantes, Nantes, FR, 2015, pp. 132.

[15] I. Khatri, H. Fukai, H. Yamaguchi, M. Sugiyama, T. Nakada, Effect of potassium fluoride post-deposition treatment on Cu(In,Ga)Se$_2$ thin films and solar cells fabricated onto sodalime glass substrates, Sol. Energy Mater. Sol. Cells, 155 (2016) 280-287.
[16] J.A. Aguiar, A. Stokes, C.-S. Jiang, T. Aoki, P.G. Kotula, M.K. Patel, B. Gorman, M. Al-Jassim, Revealing Surface Modifications of Potassium-Fluoride-Treated Cu(In,Ga)Se$_2$: A Study of Material Structure, Chemistry, and Photovoltaic Performance, Advanced Materials Interfaces, (2016) in press.
[17] P. Jackson, R. Wuerz, D. Hariskos, E. Lotter, W. Witte, M. Powalla, Effects of heavy alkali elements in Cu(In,Ga)Se$_2$ solar cells with efficiencies up to 22.6%, physica status solidi (RRL) – Rapid Research Letters, (2016) in press.
[18] T.M. Friedlmeier, P. Jackson, A. Bauer, D. Hariskos, O. Kiowski, R. Menner, R. Wuerz, M. Powalla, High-efficiency Cu(In,Ga)Se$_2$ solar cells, Thin Solid Films, (2016) in press.
[19] S.A. Jensen, S. Glynn, A. Kanevce, P. Dippo, J.V. Li, D.H. Levi, D. Kuciauskas, Beneficial effect of post-deposition treatment in high-efficiency Cu(In,Ga)Se$_2$ solar cells through reduced potential fluctuations, J. Appl. Phys., 120 (2016) 063106-063107.
[20] D. Shin, J. Kim, T. Gershon, R. Mankad, M. Hopstaken, S. Guha, B.T. Ahn, B. Shin, Effects of the incorporation of alkali elements on Cu(In,Ga)Se$_2$ thin film solar cells, Sol. Energy Mater. Sol. Cells, 157 (2016) 695-702.
[21] M. Mezher, Novel pathways to high-efficiency chalcopyrite photovoltaic devices: A spectroscopic investigation of alternative buffer layers and alkali-treated absorbers, in: Department of Chemistry and Biochemistry, University of Nevada, Las Vegas, Las Vegas, NV, 2016, pp. 164.
[22] NREL, Best Research-Cell Efficiencies, in, NCPV, Golden, CO, 2016.
[23] D. Hariskos, P. Jackson, W. Hempel, S. Paetel, S. Spiering, R. Menner, W. Wischmann, M. Powalla, Method for a high-rate solution deposition of Zn(O,S) buffer layer for high efficiency Cu(In,Ga)Se$_2$-based solar cells, in: 43rd IEEE Photovoltaic Specialists Conference, Portland, OR, 2016, pp. 1-6.
[24] M.A. Contreras, B. Egaas, P. Dippo, J. Webb, J. Granata, K. Ramanathan, S. Asher, A. Swartzlander, R. Noufi, On the role of Na and modifications to Cu(In,Ga)Se$_2$ absorber materials using thin-MF (M=Na, K, Cs) precursor layers, in: Photovoltaic Specialists Conference, 1997., Conference Record of the Twenty-Sixth IEEE, 1997, pp. 359-362.
[25] G.S. Jeong, E.S. Cha, S.H. Moon, B.T. Ahn, Effect of KF Treatment of Cu(In,Ga)Se$_2$ Thin Films on the Photovoltaic Properties of CIGS Solar Cells, Current Photovoltaic Research, 3 (2015) 65-70.
[26] Y.-S. Son, W.M. Kim, J.-K. Park, J.-h. Jeong, KF Post Deposition Treatment Process of Cu(In,Ga)Se$_2$ Thin Film Effect of the Na Element Present in the Solar Cell Performance, Current Photovoltaic Research, 3 (2015) 130-134.
[27] J.M. Raguse, C.P. Muzzillo, J.R. Sites, L. Mansfield, Effects of Sodium and Potassium on the Photovoltaic Performance of CIGS Solar Cells, Journal of Photovoltaics, (2016) in press.
[28] Z.-K. Yuan, S. Chen, Y. Xie, J.-S. Park, H. Xiang, X.-G. Gong, S.-H. Wei, Na-Diffusion Enhanced p-type Conductivity in Cu(In,Ga)Se$_2$: A New Mechanism for Efficient Doping in Semiconductors, Advanced Energy Materials, in press (2016).
[29] E. Ghorbani, J. Kiss, H. Mirhosseini, T.D. Kühne, C. Felser, Hybrid functional investigation of the incorporation of sodium and potassium in CuInSe$_2$ and Cu$_2$ZnSnSe$_4$,

in: 31st European Photovoltaic Solar Energy Conference and Exhibition, Hamburg, Germany, 2015, pp. 1271-1273.
[30] E. Ghorbani, J. Kiss, H. Mirhosseini, G. Roma, M. Schmidt, J. Windeln, T.D. Kühne, C. Felser, Hybrid-Functional Calculations on the Incorporation of Na and K Impurities into the $CuInSe_2$ and $CuIn_5Se_8$ Solar-Cell Materials, The Journal of Physical Chemistry C, 119 (2015) 25197-25203.
[31] C.S. Jiang, B. To, S. Glynn, H. Mahabaduge, T. Barnes, M.M. Al-Jassim, Recent progress in nanoelectrical characterizations of CdTe and $Cu(In, Ga)Se_2$, in: 2016 IEEE 43rd Photovoltaic Specialists Conference (PVSC), 2016, pp. 3675-3680.
[32] S. Karki, P. Paul, G. Rajan, T. Ashrafee, K. Aryal, P. Pradhan, R.W. Collins, A.A. Rockett, T.J. Grassman, S.A. Ringel, A.R. Arehart, S. Marsillac, In-situ and Ex-situ Characterizations of CIGS Solar Cells with KF Post Deposition Treatment, in: 43rd IEEE Photovoltaic Specialists Conference, Portland, OR, 2016, pp. 1-6.
[33] C.P. Muzzillo, H.M. Tong, T.J. Anderson, The effect of Na on Cu-K-In-Se thin film growth, Journal of Crystal Growth, (2016) in press.
[34] E. Handick, P. Reinhard, R.G. Wilks, F. Pianezzi, R. Félix, M. Gorgoi, T. Kunze, S. Buecheler, A.N. Tiwari, M. Bär, NaF/KF Post-Deposition Treatments and their Influence on the Structure of $Cu(In,Ga)Se_2$ Absorber Surfaces, in: 43rd IEEE Photovoltaic Specialists Conference, Portland, OR, 2016, pp. 1-5.
[35] E. Handick, P. Reinhard, J.-H. Alsmeier, L. Köhler, F. Pianezzi, S. Krause, M. Gorgoi, E. Ikenaga, N. Koch, R.G. Wilks, S. Buecheler, A.N. Tiwari, M. Baer, Potassium post-deposition treatment-induced band gap widening at $Cu(In,Ga)Se_2$ surfaces – Reason for performance leap?, ACS Appl. Mater. Interfaces, 7 (2015) 27414-27420.
[36] B. Umsur, W. Calvet, A. Steigert, I. Lauermann, M. Gorgoi, K. Prietzel, D. Greiner, C.A. Kaufmann, T. Unold, M. Lux-Steiner, Investigation of the potassium fluoride post deposition treatment on the CIGSe/CdS interface using hard x-ray photoemission spectroscopy - a comparative study, Phys. Chem. Chem. Phys., (2016).
[37] P. Pistor, D. Greiner, C.A. Kaufmann, S. Brunken, M. Gorgoi, A. Steigert, W. Calvet, I. Lauermann, R. Klenk, T. Unold, M.-C. Lux-Steiner, Experimental indication for band gap widening of chalcopyrite solar cell absorbers after potassium fluoride treatment, Appl. Phys. Lett., 105 (2014) 063901-063904.
[38] T. Maeda, A. Kawabata, T. Wada, First-principles study on alkali-metal effect of Li, Na, and K in $CuInSe_2$ and $CuGaSe_2$, Jpn. J. Appl. Phys., 54 (2015) 08KC20.
[39] D. Sommer, D. Mutter, S.T. Dunham, First-principles calculations of Na and K impurities in $CuInSe_2$ and their effect on Cd incorporation, in: 2016 IEEE 43rd Photovoltaic Specialists Conference (PVSC), 2016, pp. 2274-2278.
[40] T.M. Friedlmeier, P. Jackson, D. Kreikemeyer-Lorenzo, D. Hauschild, O. Kiowski, D. Hariskos, L. Weinhardt, C. Heske, M. Powalla, A Closer Look at Initial CdS Growth on High-Efficiency $Cu(In,Ga)Se_2$ Absorbers Using Surface-Sensitive Methods, in: 43rd IEEE Photovoltaic Specialists Conference, Portland, OR, 2016, pp. 1-5.
[41] T. Lepetit, S. Harel, L. Arzel, G. Ouvrard, N. Barreau, Coevaporated $KInSe_2$: A Fast Alternative to KF Postdeposition Treatment in High-Efficiency $Cu(In,Ga)Se_2$ Thin Film Solar Cells, IEEE J. Photovoltaics, PP (2016) 1-5.
[42] P. Reinhard, B. Bissig, F. Pianezzi, H. Hagendorfer, G. Sozzi, R. Menozzi, C. Gretener, S. Nishiwaki, S. Buecheler, A.N. Tiwari, Alkali-Templated Surface

Nanopatterning of Chalcogenide Thin Films: A Novel Approach Toward Solar Cells with Enhanced Efficiency, Nano Lett., 15 (2015) 3334-3340.
[43] T. Lepetit, S. Harel, L. Arzel, G. Ouvrard, N. Barreau, Co-evaporated $KInSe_2$: a fast alternative to KF post-deposition treatment in high efficiency $Cu(In,Ga)Se_2$ thin film solar cells, in: 43rd IEEE Photovoltaic Specialists Conference, Portland, OR, 2016, pp. 1-4.
[44] R. Wuerz, A. Eicke, F. Kessler, S. Paetel, S. Efimenko, C. Schlegel, CIGS thin-film solar cells and modules on enamelled steel substrates, Sol. Energy Mater. Sol. Cells, 100 (2012) 132-137.
[45] S. Ishizuka, A. Yamada, P.J. Fons, H. Shibata, S. Niki, Interfacial Alkali Diffusion Control in Chalcopyrite Thin-Film Solar Cells, ACS Appl. Mater. Interfaces, 6 (2014) 14123-14130.
[46] N. Barreau, P. Zabierowski, L. Arzel, M. Igalson, K. Macielak, A. Urbaniak, T. Lepetit, T. Painchaud, A. Dönmez, J. Kessler, Influence of post-deposition selenium supply on $Cu(In,Ga)Se_2$-based solar cell properties, Thin Solid Films, 582 (2015) 43-46.
[47] Y.M. Shin, C.S. Lee, D.H. Shin, H.S. Kwon, B.G. Park, B.T. Ahn, Surface modification of CIGS film by annealing and its effect on the band structure and photovoltaic properties of CIGS solar cells, Curr. Appl Phys., 15 (2015) 18-24.
[48] M. Theelen, V. Hans, N. Barreau, H. Steijvers, Z. Vroon, M. Zeman, The impact of alkali elements on the degradation of CIGS solar cells, Prog. Photovolt: Res. Appl., 23 (2015) 537-545.
[49] C.P. Muzzillo, L.M. Mansfield, K. Ramanathan, T.J. Anderson, Properties of $Cu_{1-x}K_xInSe_2$ alloys, J. Mat. Sci., 51 (2016) 6812-6823.
[50] C.P. Muzzillo, Chalcopyrites for Solar Cells: Chemical Vapor Deposition, Selenization, and Alloying, in: Chemical Engineering, University of Florida, Gainesville, FL, 2015, pp. 458.
[51] C.P. Muzzillo, H.M. Tong, T.J. Anderson, The effect of temperature on Cu-K-In-Se thin films, Chem. Mater., sumitted (2016).
[52] C.P. Muzzillo, J.V. Li, L.M. Mansfield, K. Ramanathan, T.J. Anderson, Surface and bulk effects of K in highly efficient $Cu_{1-x}K_xInSe_2$ solar cells, Advanced Energy Materials, in press (2016).
[53] C.P. Muzzillo, L.M. Mansfield, C. Dehart, K. Bowers, R.C. Reedy, B. To, R. Noufi, K. Ramanathan, T.J. Anderson, The effect of Ga content on the selenization of co-evaporated CuGa/In films and their photovoltaic performance, in: Photovoltaic Specialist Conference (PVSC), 2014 IEEE 40th, Denver, CO, 2014, pp. 1649-1654.
[54] C.P. Muzzillo, L.M. Mansfield, C. DeHart, K. Bowers, R.C. Reedy, B. To, K. Ramanathan, T.J. Anderson, Differences between CuGa/In and Cu/Ga/In films for selenization, in: Photovoltaic Specialist Conference (PVSC), 2015 IEEE 42nd, New Orleans, LA, 2015, pp. 1-6.
[55] Z.-L. Liu, S.-H. Lin, C.-H. Lu, Preparation and Characterization of Copper Gallium Diselenide Thin Films from the Solgel-Derived Precursors, Int. J. Appl. Ceram. Technol., 10 (2013) 986-993.
[56] L. Zhang, Q. He, C.-M. Xu, Y.-M. Xue, C.-J. Li, Y. Sun, The effect of composition on structural and electronic properties in polycrystalline CuGaSe 2 thin film, Chin. Phys. B, 17 (2008) 3138-3105.

[57] H. Rodriguez-Alvarez, N. Barreau, C.A. Kaufmann, A. Weber, M. Klaus, T. Painchaud, H.W. Schock, R. Mainz, Recrystallization of Cu(In,Ga)Se$_2$ thin films studied by X-ray diffraction, Acta Mater., 61 (2013) 4347-4353.

[58] J. Kim, T. Hughbanks, Synthesis and Structures of Ternary Chalcogenides of Aluminum and Gallium with Stacking Faults: KMQ$_2$ (M=Al, Ga; Q=Se, Te), J. Solid State Chem., 149 (2000) 242-251.

[59] H.A. Al-Thani, M.M. Abdullah, F.S. Hasoon, XPS investigation of surface secondary phase segregation in CIGS thin film, in: 2011 37th IEEE Photovoltaic Specialists Conference, 2011, pp. 000315-000319.

[60] M.R. Balboul, U. Rau, G. Bilger, M. Schmidt, H.W. Schock, J.H. Werner, Control of secondary phase segregations during CuGaSe$_2$ thin-film growth, Journal of Vacuum Science & Technology A, 20 (2002) 1247-1253.

[61] V. Nadenau, G. Lippold, U. Rau, H.W. Schock, Sodium induced secondary phase segregations in CuGaSe$_2$ thin films, Journal of Crystal Growth, 233 (2001) 13-21.

[62] X. Song, R. Caballero, R. Félix, D. Gerlach, C.A. Kaufmann, H.-W. Schock, R.G. Wilks, M. Bär, Na incorporation into Cu(In,Ga)Se$_2$ thin-film solar cell absorbers deposited on polyimide: Impact on the chemical and electronic surface structure, J. Appl. Phys., 111 (2012) 034903.

[63] B.J. Stanbery, E.S. Lambers, T.J. Anderson, XPS studies of sodium compound formation and surface segregation in CIGS thin films, in: Conference Record of the Twenty Sixth IEEE Photovoltaic Specialists Conference - 1997, 1997, pp. 499-502.

[64] S.-H. Wei, S.B. Zhang, A. Zunger, Effects of Na on the electrical and structural properties of CuInSe$_2$, J. Appl. Phys., 85 (1999) 7214-7218.

[65] M.I. Alonso, K. Wakita, J. Pascual, M. Garriga, N. Yamamoto, Optical functions and electronic structure of CuInSe$_2$, CuGaSe$_2$, CuInS$_2$, and CuGaS$_2$, Physical Review B, 63 (2001) 075203.

[66] H.-W. Ma, Guo, M.-S. Wang, GuoZhou, S.-H. Lin, Z.-C. Dong, J.-S. Huang, K$_2$MM'$_3$Se$_6$ (M = Cu, Ag; M' = Ga, In), A New Series of Metal Chalcogenides with Chain−Sublayer−Chain Slabs: $_\infty^1$[M'Se$_4$]−$_\infty^2$[(MSe$_4$)(M'Se$_4$)]−$_\infty^1$[M'Se$_4$], Inorg. Chem., 42 (2003) 1366-1370.

[67] K. Feng, D. Mei, L. Bai, Z. Lin, J. Yao, Y. Wu, Synthesis, structure, physical properties, and electronic structure of KGaSe$_2$, Solid State Sci., 14 (2012) 1152-1156.

[68] Z.Z. Kish, V.B. Lazarev, E.Y. Peresh, E.E. Semrad, Compounds in In$_2$Se$_3$-K$_2$Se, Neorg. Mater., 24 (1988) 1602-1605.

[69] P. Wang, X.-Y. Huang, Y.-L. Liu, Y.-G. Wei, J. Li, H.-Y. Guo, Solid state synthesis at intermediate temperature and structural characterization of KInSe$_2$, Acta Chim. Sinica, 58 (2000) 1005-1008.

[70] B. Eisenmann, A. Hofmann, Crystal structure of pentapotassium tetraselenidogallate(III), K$_5$GaSe$_4$, Z. Kristallogr., 197 (1991) 163-164.

[71] B. Eisenmann, A. Hofmann, Crystal structure of hexapotassium di-μ-selenido-bis(diselenidogallate), K$_6$Ga$_2$Se$_6$, Z. Kristallogr., 197 (1991) 153-154.